\documentclass[10pt,journal,compsoc]{IEEEtran}
\ifCLASSOPTIONcompsoc
  \usepackage[nocompress]{cite}
\else
  \usepackage{cite}
\fi
\ifCLASSINFOpdf
  \usepackage[pdftex]{graphicx}
  \graphicspath{{png/}{../png/}}
  \DeclareGraphicsExtensions{.pdf,.jpeg,.png}
\else
  \usepackage[dvips]{graphicx}
  \graphicspath{{png/}{../png/}}
  \DeclareGraphicsExtensions{.png}
\fi
\usepackage{amsmath}
\usepackage{algorithm}
\usepackage{algorithmic}

\usepackage{array}

\ifCLASSOPTIONcompsoc
 \usepackage[caption=false,font=footnotesize,labelfont=sf,textfont=sf]{subfig}
\else
 \usepackage[caption=false,font=footnotesize]{subfig}
\fi
\usepackage{fixltx2e}
\usepackage{dblfloatfix}

\ifCLASSOPTIONcaptionsoff
 \usepackage[nomarkers]{endfloat}
\let\MYoriglatexcaption\caption
\renewcommand{\caption}[2][\relax]{\MYoriglatexcaption[#2]{#2}}
\fi
\usepackage{url}

\usepackage{booktabs}
\usepackage{bbm}
 \usepackage{threeparttable}

\begin{document}
%
\title{Hashing-Based Distributed Clustering\\ for Massive High-Dimensional Data}
%
%
%
%

\author{Yifeng~Xiao,
        Jiang~Xue,~\IEEEmembership{Senior Member,~IEEE,}
        and~Deyu~Meng
\IEEEcompsocitemizethanks{\IEEEcompsocthanksitem Yifeng Xiao are with the Xi'an Jiaotong University, Shaanxi 710049, China (e-mail: yifengxiao.xjtu@gmail.com).\protect\\
 Jiang Xue and Deyu Meng are with the School of Mathematics
and Statistics, Xi’an Jiaotong University, Xi’an 710049, China, and
also with the Pengcheng Laboratory, Shenzhen 518000, China (e-mail:
x.jiang@xjtu.edu.cn; dymeng@xjtu.edu.cn). Jiang Xue is the corresponding author.}
}

%
%

\markboth{Hashing-Based Distributed Clustering}%
{Yifeng \MakeLowercase{\textit{et al.}}: Hashing-Based Distributed Clustering}
%



\IEEEtitleabstractindextext{%
\begin{abstract}
Clustering analysis is of substantial significance for data mining. The properties of big data raise higher demand for more efficient and economical distributed clustering methods. However, existing distributed clustering methods mainly focus on the size of data 
but ignore possible problems caused by data dimension. To solve this problem, we propose a new distributed algorithm, referred to as Hashing-Based Distributed Clustering 
 (HBDC). Motivated by the outstanding performance of hashing methods for nearest neighbor searching, this algorithm applies the learning-to-hash technique to the clustering problem, which possesses incomparable advantages for data storage, transmission and computation. Following a global-sub-site paradigm, the HBDC consists of distributed training of hashing network and spectral clustering for hash codes at the global site. The sub-sites use the learnable network as a hash function to convert massive HD original data into a small number of hash codes, and send them to the global site for final clustering. In addition, a sample-selection method and slight network structures are designed to accelerate the convergence of the hash network. We also analyze the transmission cost of HBDC, including the upper bound. Our experiments on synthetic and real datasets illustrate the superiority of HBDC compared with existing state-of-the-art algorithms. 

\end{abstract}

}

\maketitle

\IEEEdisplaynontitleabstractindextext

%
\IEEEpeerreviewmaketitle

\IEEEraisesectionheading{\section{Introduction}\label{sec:introduction}}

%
%
%
%
\IEEEPARstart{C}{lustring} is a classical unsupervised technique widely used to discover the latent structure within a large dataset.
Specifically, clustering aims to classify samples in one dataset into several
clusters by their distribution features and maximize the similarity of the samples
in one cluster while minimizing the samples' similarity between different clusters. There are various clustering algorithms that can distinguish clusters effectively
on different kinds of datasets. However, with the arrival of the big-data era, the change in data properties brings new challenges. Real-world data are usually 
generated and stored in distributed machines\cite{2004DBDC}, which cannot meet the requirements
of centralized clustering. Whereas collecting all the data into a central computer is 
nearly impossible because of the unaffordable transmission cost and privacy concerns. Meanwhile,
the high dimension of big data also results in the curse of dimensionality\cite{2015Dynamic} and the rising of computational complexity. It may be impossible to process
massive high-dimensional data in a single computer. Therefore, how to process and cluster data
in a distributed scenario 
is an inevitable problem.

Some distributed clustering algorithms have been proposed to solve this problem in a global-sub-site paradigm consisting of one global site and several sub-sites. In order to reduce the overhead, existing distributed clustering methods try to model local data at the sub-site and only send the model parameters to the global site, where an overall model is generated by merging all local models and broadcast to 
all the sub-sites. To be specific, there are basically four groups: 
density-based methods\cite{10.1007/978-3-319-33747-0_8}, partition-based methods\cite{0Data},grid-based methods\cite{2006Survey} and model-based methods\cite{2020Local}. Density-based methods search for high-density regions as clusters and collect some core samples as cluster parameters. Partition-based methods execute preliminary local clustering at sub-sites
and then upload the centers and their weights (the number of samples belonging to the cluster).
Grid-based methods divide sample space with finite grids, and the samples are grouped by their grid 
neighborhood relationship. Model-based methods always pre-define a statistical
model and try to fit local data by parameter estimation. These proposed methods can process massive distributed data and have achieved remarkable clustering performance.

Recently, the data in empirical applications, such as 
image retrieval\cite{2019Clustering} and text analysis\cite{2005Text}, usually show the characteristics of high dimension (HD). As a result of the explosive growth of data volume in the real world, several problems are involved. 
The first one is the curse of dimensionality\cite{2015Dynamic}. Points in HD space are usually of sparsity so that the variance of estimated density is inflated as the dimensionality of data increases to hundreds or even more, which results in a degradation of clustering performance. In addition, the high-density regions searched in HD space are also inaccuracy for density-based methods. 
Second, the shape of clusters becomes irregular and complicated, while partition-based methods are based on some centralized convex clustering methods\cite{2007Ensemble}, which depends on an assumption that clusters are convex. However, it is usually not true for HD space situations, so the performance cannot be guaranteed.
For the grid-based methods, the total number of grids shows exponential growth with dimension increasing\cite{amini2011study}. Correspondingly, the computational cost becomes unaffordable when processing HD data. The last problem is that parameter estimation for HD statistical models often costs enormous computational resources, so most model-based methods have to make a compromise between the cost and accuracy\cite{2014Model}. Overall, distributed clustering for HD data remains to be solved.

Nearest neighbor (NN) searching, which aims to find the point with the smallest distance to a target query in a given set\cite{Hashing}, is a basic step in a wide range of tasks and can also be regarded as a fundamental sub-problem of clustering. Even when it comes to HD space, the computational cost can be limited by approximate methods, and the results can still meet requirements in most cases.
For the NN problem, hashing methods show incomparable advantages in terms of the efficiency of storage and calculation\cite{2021Learning} and thus gain great popularity. It aims to use a hash function to map the HD original data into binary hash codes, and the similar objects in the HD space are supposed to be converted to the hash codes with smaller distances. The calculation of hamming distance and the storage of hash codes obviously cost much lower than those of the original data. Traditional hashing method, which is called local density hashing(LSH)\cite{Charikar2002Similarity},  builds hash maps to map data into a limited number of hash buckets, and makes similar objects map into the same buckets with higher possibility. 
However, to improve the recall rate\cite{Hashing}, these LSH methods have to build many hash maps so that their applications remain difficult when processing large datasets. Since the hash maps in LSH are independent of data, a hash function learned from the data automatically may be a more suitable choice for building high-quality codes. The development of machine learning and neural networks makes it possible. Learning to hash, a combination of hashing and neural network, gets wide attention because of its outstanding performance and capability to process large and complex datasets. 

Motivated by the superiority of learning to hash in NN searching and the similarity between NN searching and clustering, a novel  distributed clustering approach for HD data, referred to as Hashing-Based Distributed Clustering (HBDC) algorithm, is proposed in this paper. This algorithm is designed for a global-sub-site paradigm which is composed of one global site and several sub-sites. First, an initial neural network as a hash function is broadcast from the global site to all the sub-sites, and a mini-batch of samples in each sub-site is selected for local training of the hash function. After the local training, the global site will collect renewed parameters of the hash function from sub-sites and do integration. This distributed machine-learning process lasts several iterations until the hash function is capable enough to generate representative hash codes. Second, all the sub-sites use this hash function to map local data into hash codes and send the hash codes to the global site for final clustering. Massive HD data are converted into a limited number of hash codes, and thus the transmission and computational cost can be reduced sharply. The contribution of this work can be summarized as follows.
\begin{itemize}
    \item [1)]
    An algorithm HBDC for distributed clustering is proposed to tackle massive HD data.
    \item[2)]
    A distance-driven self-supervised training method for hash functions focusing on clustering is proposed and implemented to get reliable hash functions efficiently.
    \item[3)]
    An analysis is presented to illustrate that HBDC is cost-effective for HD data in distributed cases.
\end{itemize}

The rest of the paper is organized as follows. In Section \uppercase\expandafter{\romannumeral2}, some related works are introduced. Section \uppercase\expandafter{\romannumeral3} gives some symbols and notations. Section \uppercase\expandafter{\romannumeral4} presents the proposed distributed clustering method HBDC and Section \uppercase\expandafter{\romannumeral5} shows some experimental results to justify the efficacy of the proposed HBDC. Finally, the conclusion is in Section \uppercase\expandafter{\romannumeral6}.

\section{RELATED WORKS}

\subsection{Distributed Clustering}
The aim of distributed clustering methods is to make clustering analysis of data discretely stored in distributed different sites without direct transmission of the original data. Therefore, the algorithm design is substantially influenced by the topological properties of site networks. Generally, there are two network paradigms that are mainly researched: peer-to-peer (P2P) networks and global-sub-site networks\cite{2014Models}. The main difference between them is whether there is a global site that has access to all sub-sites to collect necessary data for clustering. In the P2P paradigm, there is not such a global site, and each sub-site can only communicates with a few other sites for local clustering\cite{2014Models}. By contrast, the global-sub-site paradigm has a global site acting as a central server, and there is no direct communication between sub-sites. In this paper, only the global-sub-site paradigm is discussed.

In \cite{2004DBDC}, Januzaj \textit{et al}. proposed a DBSCAN-based distributed clustering method called DBDC. In the first step, it clusters data in the sub-sites locally by DBSCAN. In each local cluster, some core samples are selected and associated with their neighborhood parameters. These samples and parameters are sent to the global site and participate in global DBSCAN clustering to generate final clusters. Though DBDC can find latent clusters with arbitrary shapes, it also inherits the downside of DBSCAN, the sensitivity to input parameters, which means that DBDC method can only perform well when receiving proper parameters. For that matter, \cite{2008Local} improved DBDC and made good progress in clustering performance. On the other hand, in \cite{1999A}, Xu \textit{et al}. used DBSCAN to process centralized clustering parallel by dividing dataset at global site into some clusters, and then used distributed machines to handle each cluster parallel.
Overall, these density-based methods can find clusters with complex shapes but are highly dependent on the input parameters.

In \cite{2005Privacy}, Jagannathan and Wright extended the basic clustering algorithm k-means into the distributed case. Meanwhile, privacy protection was also considered by avoiding exposure of original private data. It performs well when the data in different sites represent different attributes of common entities. Ji and Ling\cite{2007Ensemble} proposed an ensemble-learning-based clustering algorithm. In this distributed k-means clustering algorithm, the entire clustering process is divided into two steps. First, every sub-site executes local k-means clustering and summarizes the centroid points of all clusters and the numbers of points in each cluster as the weights of centroid points. Then all the centroid points and their weights are sent to the global site,  where another k-means clustering to centroid points is done. Based on the results of the two clusterings, sub-sites can get the final result. Essentially, the whole clustering process is composed of a 2-layer hierarchical clustering and both sub-site and global site take one of them. In \cite{2013Distributed}, Balcon \textit{et al}. proposed a distributed k-means clustering based on core-set of samples to process distributed data more efficiently. Jeong \textit{et al}.
\cite{2007Integration} focused on biological data and modified distributed k-means algorithm. This modified algorithm can achieve outstanding performance on enormous datasets. Overall, the foregoing methods are efficient in handling distributed clustering tasks, whereas their performance is highly limited by their basis k-means algorithm.

A model-based distributed clustering algorithm based on expectation maximum (EM) method\cite{1977Maximum} was proposed by Kriegel \textit{et al}. in 2005\cite{2005Effective}. It uses a mixture of Gaussian distributions to model local data, and the parameters of distribution functions are estimated by EM method. Then, these Gaussian mixture distributions from all sub-sites are merged at the global site and form a general global model. For the concern of privacy, a modified algorithm in \cite{2005Privacy} limits data sharing between sites to preclude possible data divulging. Another algorithm referred to LDSDC\cite{2020Local} uses subspace Gaussian model, which is flexible to data dimension, to do local clustering at sub-sites and achieve better performance on HD datasets. But the subspace Gaussian model contains parameters whose number rises dramatically with data dimension, so the cost is still high. Furthermore, sometimes the Gaussian model cannot fit datasets well, which limits its performance.
On the other hand, there are also some centralized methods aiming to directly process HD data. In \cite{1998Automatic}, R Agrawal \textit{et al}. proposed a grid-based algorithm CLIQUE. It divides feature space into isometric intervals in every dimension, and the grids which contain more than a certain threshold number of samples are regarded as dense cells. Then, adjacent dense cells are merged to constitute clusters in a bottom-up fashion. CLIQUE is highly dependent on proper input parameters, including threshold and grid size.  Another up-bottom grid-based method WFC\cite{2022Massive} overcomes the drawbacks of CLIQUE by using a Weber–Fechner-Law-inspired approach to calculate different scales and obtain multi-scale clustering results. However, these centralized clustering methods cannot be extended to distributed cases while maintaining their efficacy and economy.

\subsection{LEARNING TO HASH}
To solve the nearest neighbor searching problem, researchers have made a wide range of endeavors to learn compact hash functions from data to preserve the distance order in original space. Y Gong \textit{et al}. proposed AQBC\cite{2012Angular}, which uses cosine similarity between two samples as the measure and maps vectors into the nearest binary hypercube. In \cite{2017Fast}, Jie Gui \textit{et al}. regressed semantic labels of samples to hash codes in a supervised manner and then optimized the hash codes. Many further types of research \cite{1999Similarity}\cite{2011LDAHash}\cite{2021Probability} have been proposed to achieve better performance. However, limited by the ability of representation, these simple hashing methods are not scalable for massive and complex datasets.

With the development of deep learning, more and more researchers are trying to apply deep neural networks to hashing algorithms, called deep hashing (DH). DH uses hashing neural networks as hash functions to map data into hash codes in an end-to-end manner. Generally, the DH methods can be divided into supervised DH and unsupervised DH, and researchers mainly focus on the supervised methods. The core issue of current supervised methods is how to measure the similarities of samples. There are three categories generally: point-wise methods, pairwise methods and ranking-based methods. CQS\cite{9156349} employs a point-wise method that uses label information directly instead of similarity information. First, it generates some central hash codes with different class labels and then enforces the outputs of networks to approach the hash centers with the same labels. In \cite{2016Deep}, H Liu \textit{et al}. proposed a pair-wised method DSH. It uses labels representing whether a pair of samples are semantically similar to train a hashing network by minimizing the hamming distances of the hash codes with similar labels and maximizing those with different labels. A network consisting of three convolutional layers and two following fully-connected layers is designed as the hash function of DSH. H Lai \textit{et al}. proposed DNNH\cite{7298947} based on the ranking method. It groups samples into triplets and encourages hash codes of more similar samples to be closer. 

The supervised methods are the basis of the unsupervised ones. To compensate for the loss of label information, most unsupervised methods utilize pre-trained networks to obtain semantic information, and convert the problem into a supervised one. However, some works still chose to use data themselves for training without generating semantic information. For example, AETBH\cite{9157770} and BinGAN\cite{10.5555/3327144.3327278} introduced self-supervised techniques into unsupervised deep hashing and used auto-encoder and generative adversarial networks as hash functions, respectively. In \cite{9711357}, YK Jang and NI Cho involved contrastive learning in the training of hash functions. The contrastive-learning method uses one sample to generate more inputs by adding noise randomly. The inputs deriving from the same samples should be mapped to closer hash codes. Although these unsupervised methods can generate high-quality hash codes and achieve promising performance, the computational cost is usually very high because of the heavy structure of neural networks. In distributed cases, it will result in unacceptable transmission cost, so more slight networks should be chosen. The HBDC proposed in this paper provides a feasible solution.

\begin{figure*}[t]
    \centering
    \includegraphics{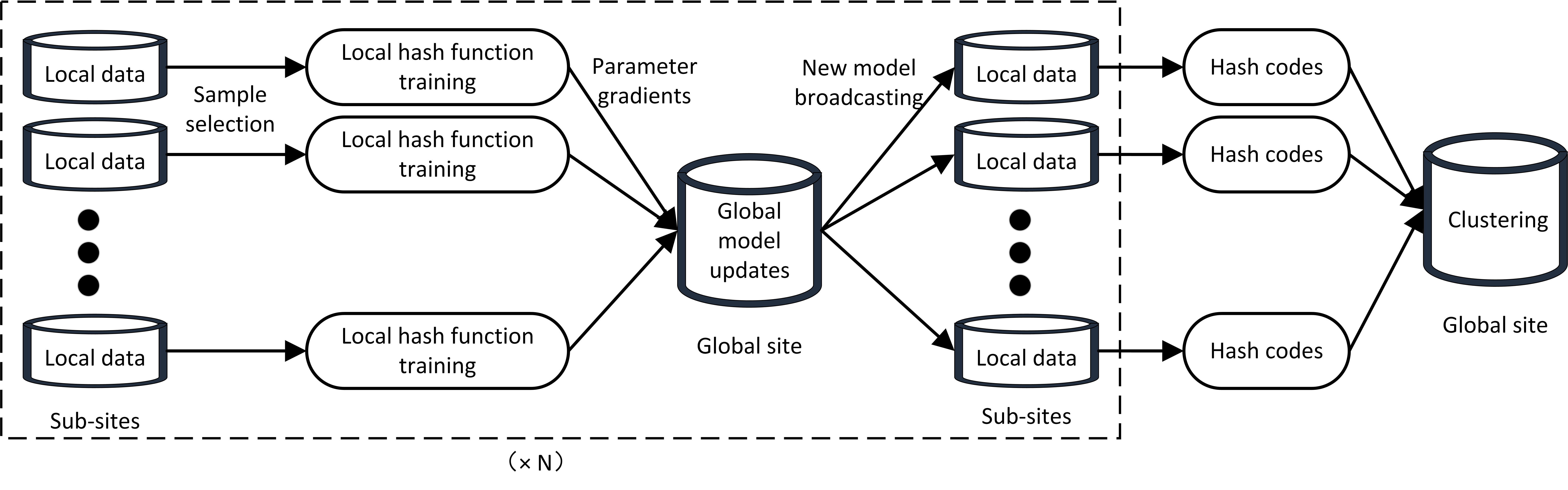}
    \caption{The progress of the HBDC algorithm. The distributed training of hashing function lasts $N$ iterations.}
    \label{fig:1}
\end{figure*}
\section{SYMBOLS and NOTATIONS}
Some notations are defined as follows. Let lower- (e.g., $a$) and upper-case (e.g., $A$) letters denote scales. $\mathcal{L}$ denotes the loss function. Boldface lower- (e.g., $\textbf{a}$) and upper-case letters (e.g., $\textbf{A}$) denote vectors and matrices. Calligraphic letters (e.g., $\mathcal{A}$) donate sets. $\textbf{a}_{i.}$ and $\textbf{a}_{.j}$ are the $i$th row and the $j$th column of a matrix.$A$. We use $\Vert \cdot \Vert_1$ and $\Vert \cdot \Vert_2$ to represent the 1-norm and 2-norm of vectors. $|\cdot|$ donates the cardinality of a set. $x_i(\textbf{X})$ , $h_i(\textbf{H})$ and $b_i(\textbf{B})$ donate the inputs, outputs of neural networks and the output hash codes, respectively (in matrix form). $L$ donates the length of hash codes. $\Psi(\cdot)$ donates the neural networks, and $\Theta$ donates the set of their neural parameters.
A graph $G$ consists of $m$ vertices (denoted by $V = {v_1, v_2, ..., v_m}$) and some edges (denoted by $E = {(v_i, v_j)| v_i, v_j \in V}$).

\section{HASHING-BASED DISTRIBUTED CLUSTERING}

In this section, the HBDC algorithm is presented. Consider a situation that the whole system consists of a global site and $M$ sub-sites. The total data that need to be clustered are stored in $W$ sub-sites dispersedly 
$\mathcal{X} = \cup_{m=1}^{M}\mathcal{X}_m$. The local data in the $m$th sub-site are $s$-dimensional vectors 
$\mathcal{X}_m = \{\textbf{x}_i^m\in\mathbbm{R}^D\}_{i=1}^{V_m}$, where ${V_m}$ donates the number of samples stored in the $m$th sub-site. As presented in Fig. \ref{fig:1}, HBDC aims to cluster data in three steps:
\begin{itemize}
    \item[1)] The global site broadcasts an initiative hashing network to all the sub-site. In each sub-site, a batch of samples is selected for the local training of the hashing network.
    \item[2)] The parameter gradients of the network are sent to the global site, and then the global site upgrades the hashing network by merging all gradients. The first two steps repeat $N$ times until the convergence.
    \item[3)] The sub-sites use the hashing network to convert local data into hash codes, and then the codes and their degrees (i.e., the number of samples each code represents) are sent to the global site. A graph is generated with hash codes as its vertices. Thus the clustering problem can be converted into a graph-cutting problem. Based on graph theory, the graph is cut by a normalized-cut method.  
\end{itemize}

The following subsections discuss the distributed training of hash function and the hash code clustering in terms of details and analysis.

\subsection{Distributed Training of Hash Function}
Unlike existing DH methods, the overall hash function for distributed clustering cannot be trained in a single machine, which means that the overhead during the training process must be considered. The transmission cost for training can be calculated as:
\begin{equation}\label{eq1}
    C_t = M \times |\Theta| \times 2N \times 32 \; (bit),  
\end{equation}
where $|\Theta|$ donates the number of parameters of hashing network, $N$ donates the number of iterations in the training process and a \textit{float32} data accounts for $32$ bits. $2N$ contains data transmitted in unlink and downlink communication. For simplicity, this research does not consider quantization techniques of parameters for cost reduction\cite{zheng2020design}\cite{fed}. All network parameters are transmitted as original \textit{float32} datatype. 

To limit the cost, we need to reduce the number of network parameters and accelerate the convergence. Centralized DH methods usually use complex networks with over millions parameters to achieve promising performance, such as ResNet50\cite{xie2017aggregated}, while distributed training of these networks can result in enormous communication cost. Thus, in this paper, more slight and compact networks are used as hashing networks.
\begin{figure}[b]
    \centering
    \includegraphics{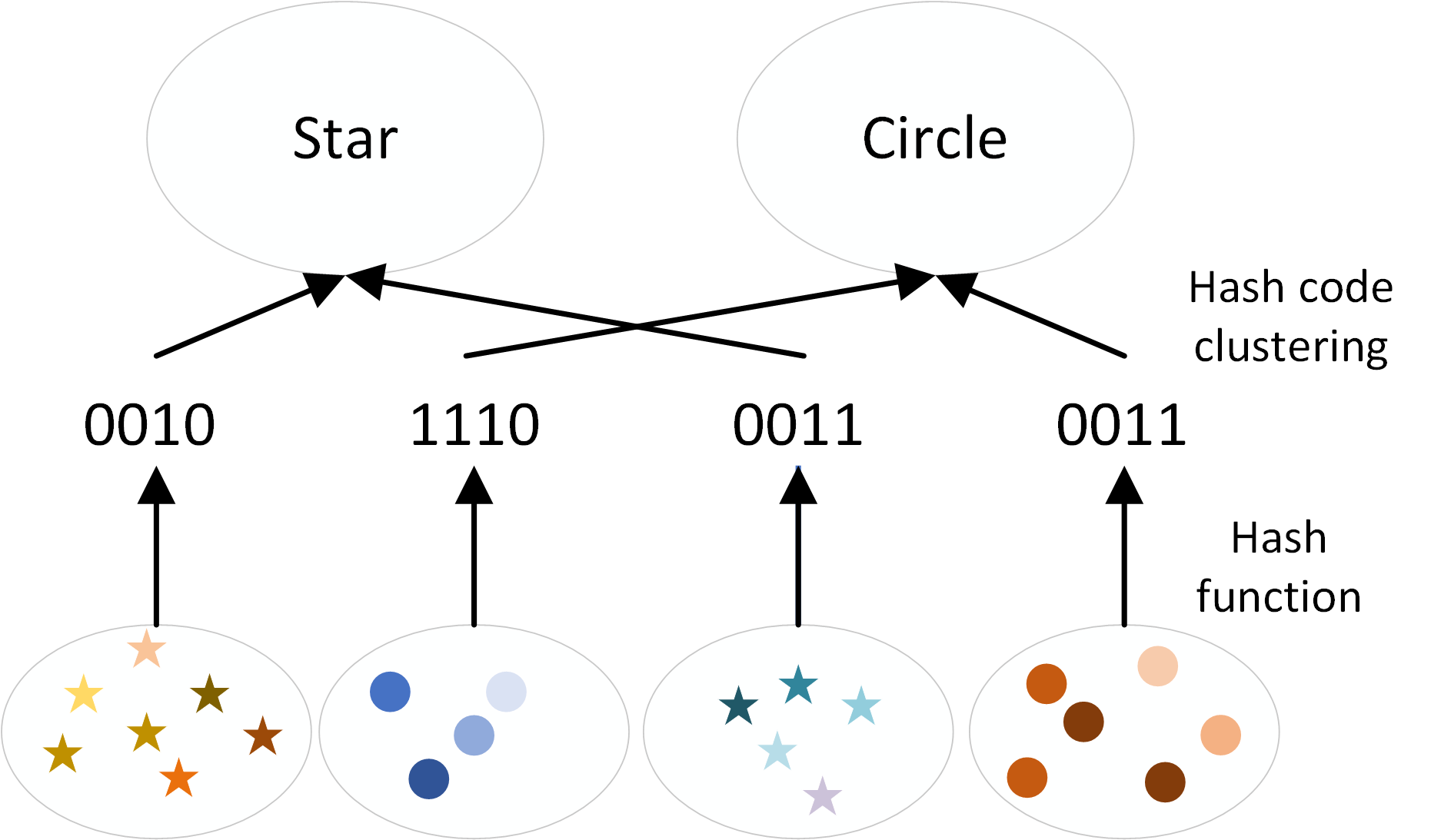}
    \caption{Illustration of the hierarchical clustering in the HBDC. The hash function is only responsible for the first step.}
    \label{fig:2}
\end{figure}

On the other hand, to compensate for the loss of representation ability caused by the network structure, the whole clustering is divided into two steps hierarchically. The hash codes represent the result of the first-step clustering, which means the hashing network does not have to be powerful enough to generate an end-to-end result.
As shown in Fig.\ref{fig:2}, the shape information is coupled with color in the hash codes, but later clustering can group the hash codes into proper categories. In more complex datasets with more features, this can be achieved by increasing the length of hash codes $L$. In this way, a part of network function can be transferred to the second step.

\begin{figure}[t]
    \centering
    \includegraphics{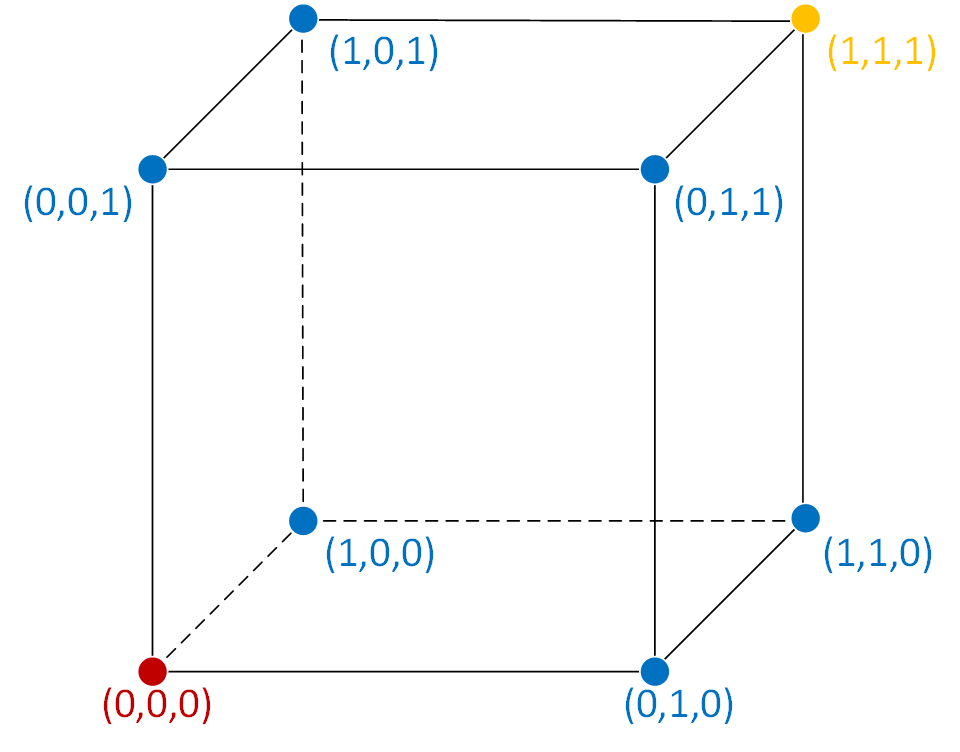}
    \caption{An example of sample selection. First, a sample with the hash code represented by the red point is chosen randomly, and then the second one is the opposite vertex colored yellow to maximize its hamming distance from the first one. The third sample is randomly chosen from the six blue points.}
    \label{fig:3}
\end{figure}

In addition, to increase the convergence rate, a filter is designed to make the samples for local training more representative. The principle of selection is to make the samples as different as possible. After randomly choosing the first sample, the second one comes from the cluster with the hash code, which has the biggest hamming distance from the first one. Following this mechanism, the $i$th sample is chosen randomly from the cluster that maximizes the sum of hamming distances from the previous $i-1$  codes until the whole batch of samples is  determined. An illustration of the mechanism of sample selection is shown in Fig.\ref{fig:3}

The local training of hashing network exploits the pair-wise method. After normalization, two samples are grouped together, and the model is trained by minimizing the difference between the distances of input space and hamming space, i.e.,
\begin{equation}\label{eq2}
\begin{split}
    &\mathcal{L}^m = \vert\lambda\Vert \textbf{x}_{i}^m - \textbf{x}_{j}^m\Vert_2 - \Vert \textbf{b}_i^m - \textbf{b}_{j}^m\Vert_1\vert e^{-\frac{\Vert \textbf{x}_{i}^m - \textbf{x}_{j}^m\Vert_2}{t}},\\
    s.t. \quad & \textbf{b}_{i}^m, \textbf{b}_{j}^m \in \{-1, 1\}^L,
\end{split}
\end{equation}
where $\textbf{x}_{i}^m$, $\textbf{x}_{j}^m$ donate the inputs in $m$th sub-site, $\textbf{b}_i^m = \Psi(\textbf{x}_{i}^m;\Theta)$ donates the hash codes and $\lambda$, $t$ are hyper-parameters. Since the maximum hamming distance between hash codes is $2L$, the inputs with distances more than $2\lambda L$ reward hash function maps them to totally different codes. Thus we can encourage the hash function to yield more kinds of hash codes to use the $L$bits to the greatest extent by adjusting the parameter $\lambda$.

Because of the gradient property of sign function, the gradients of $\Theta$ cannot be obtained by back-propagating. Thus, $\tanh({\cdot})$ is used to approximate the sign function in the training part, and problem (\ref{eq2}) is rewritten as:
\begin{equation}\label{eq3}
    \widetilde{\mathcal{L}_m} = \vert\lambda\Vert \textbf{x}_{i}^m - \textbf{x}_{j}^m\Vert_2 - \Vert \textbf{h}_i^m - \textbf{h}_{j}^m\Vert_1\vert e^{-\frac{\Vert \textbf{x}_{i}^m - \textbf{x}_{j}^m\Vert_2}{t}},
\end{equation}
where $\textbf{h}_{i}= \widetilde{\Psi}(\textbf{x}_{i};\Theta)$ donate the outputs of hashing network with the last hashing layer replaced by $\tanh({\cdot})$. This replacement is only used for upgrading neural parameters $\Theta$. After that, the sign function will be used to generate hash codes.

After the local training, the local average parameter gradients $\nabla \Theta^m_n$ are transmitted to the global site and then it can upgrade the model to the next iteration with the average:
\begin{equation}\label{eq4}
    \Theta_{n+1} = \Theta_{n} - l\sum_{m=1}^M\frac{\nabla \Theta^m_n}{M},
\end{equation}
where the parameter $l$ denote learning rate. Although we assume that all the sub-sites have the same batch size, (\ref{eq4}) can be easily extended to some cases in that different sub-sites have different batch-size, such as online clustering. Although the batch size in each sub-site may be small, the total size of all batches can be much larger to make the model upgrade more reliable. The hashing network with updated parameters is broadcast to the sub-sites to begin the next training iteration. This iteration is repeated $N$ times to get a high-quality hash function.

\subsection{Hash Code Clustering}
After training, the sub-sites use the hashing network to transform all local data into hash codes and count the number of samples corresponding to each hash code as its degree. Compared with massive original data, the number of hash codes is much smaller, and their binary format makes it much more convenient for storage, transmission and computation. The hash codes and their degrees are sent to the global site for final clustering. In this communication, the volume of transmitted data is
\begin{equation}\label{eq5}
    C_c = \sum_{m=1}^M num_m \times (32 + L) (bit),
\end{equation}
where $num_m$ donates the number of hash codes in the $m$th sub-site. $L$ is the length of hash codes, and the degrees are transmitted as the data type \textit{float32}.

Since the hash codes are actually distributed on the vertices of an $L$-dimensional cube, clustering for hash codes may not be convex. Thus, justifiably exploiting simple clustering methods on hash codes may not be reasonable. For that matter, we use the codes to generate a graph and exploit spectral clustering\cite{von2007tutorial} on it. Spectral clustering is a class of clustering algorithms working on an undirected graph. It divides the vertices into a given number of clusters by cutting the edges between them. There are different methods of graph cutting, such as \textit{Radio Cut}\cite{radiocut} and \textit{Normalized Cut}\cite{normalizedcut}
In this research, the \textit{Normalized Cut} method is chosen, which aims to minimize the following loss function:
\begin{equation}\label{eq6}
\begin{split}
    &\mathcal{L}_{NCut}(G_1,G_2,...,G_k) = \frac{1}{2}\sum_{i=1}^k\frac{W(G_i,\overline{G_i})}{vol(G_i)},\\
    s.t. \quad &\cup_{i=1}^{k}G_i = G,\\
    &G_i \cap G_j = \emptyset,
\end{split}
\end{equation}
where $G_1,G_2,...,G_k$ are sub-graphs, $vol(\cdot)$ represents the number of vertices, and $W(G_i,\overline{G_i})$ is the sum of edge weights connecting the sub-graph $G_i$ and its complementary sub-graph $\overline{G_i}$. This graph cutting can be solved by matrix analysis method, which is detailed in \cite{normalizedcut}. 
Note that the feasibility of spectral clustering is also enabled by the superiority of hash codes because calculating an adjacent matrix between millions of HD samples is time-consuming, while it is much easier when processing much fewer binary hash codes.

In the hash code clustering, the undirected graph $G$ is generated by the hash codes and their degree. The vertices of $G$ denote the hash codes. The weights of edges between codes are calculated by their hamming distances and the degrees of codes as follows:
\begin{equation}\label{eq7}
    w_{(b_i,b_j)} = \frac{d_{b_i}\times{d_{b_j}}}{hamming(b_i,b_j)}.
\end{equation}
The weight of an edge represents how strongly two vertices are connected, i.e., how possible these two vertices are in the same cluster. Thus, besides the reciprocal of hamming distance, the degrees of hash codes are also considered to generate the graph. The later graph cutting tries to cut the edges with low weights, and this is the same with density-based clustering methods, which partition clusters by low-density regions. Therefore, the input adjacent matrix for spectral clustering consists of the weights in (\ref{eq7}):  
\begin{equation}\label{eq8}
    \textbf{W}_{ij} = w_{(b_i,b_j)}.
\end{equation}

\begin{algorithm}[htb]
    \caption{Hashing-based Distributed clustering}
    \label{a1}
    \renewcommand{\algorithmicrequire}{\textbf{Input:}}
    \renewcommand{\algorithmicensure}{\textbf{Output:}}
    \begin{algorithmic}[1] 
        \REQUIRE \ 
        Dataset$\mathcal{X}$, class number $k$, number of iterations $N$, network structure $\Psi$
        \ENSURE \ 
        Clustering result ${(c_1,c_2,...,c_k)}$.
        \STATE Randomly initialize parameters $\Theta_0$.
        \WHILE{$n\leq N $}
        \STATE Broadcast $\Theta_n$ to sub-sites.
        \STATE Local hashing network training with selected samples.
        \STATE Upload local parameter gradients $\Theta_n^m$ to the global site.
        \STATE Update parameters with $\Theta_{n+1}=\Theta_{n}-l\sum_{m=1}^M\frac{\nabla \Theta^m_n}{M}$.
        \STATE $n = n + 1$.
        \ENDWHILE
        \STATE Broadcasting the latest hashing network to the sub-sites.
        \STATE Map data to hash codes and count their degrees. Send them to the global site.
        \STATE Calculate Gram matrix with $\textbf{W}_{ij} = \frac{d_{b_i}{d_{b_j}}}{hamming(b_i,b_j)}$.
        \STATE Spectral cluster on the adjacent matrix.
    \end{algorithmic}
\end{algorithm}

\subsection{Algorithm Analysis}
In summary, a whole algorithm description is presented in Algorithm \ref{a1}. It is clear that this algorithm can be modified by quick start with a pre-trained network. This can improve clustering performance, dramatically reduce the training cost, and even skip the training process. Besides, choosing suitable network structures can act as a similar role.  
Compared with existing distributed clustering algorithms, the HDBC is more robust to different kinds of datasets because of the selectivity of hashing networks.

One of the most important considerations of distributed algorithms is the transmission cost. In the HDBC, it consists of two parts: the cost for distributed training and for hash code transmission.
According to (\ref{eq1}) and (\ref{eq5}), the total cost is:
\begin{equation}\label{eq9}
    C = 32(2N+1)M|\Theta| + \sum_{m=1}^M (32 + L)num_m (bit).
\end{equation}
The first term is different from (\ref{eq1}) because the broadcasting of the latest parameters (the $9th$ step in Algorithm 
 \ref{a1}) is included. For simplicity, we do not consider extra quantization methods for transmitted data to reduce the cost. In (\ref{eq9}), the size of parameter set $|\Theta|$ and the number of hash codes $num_m$ are two key factors influencing the cost mostly. No matter what kind of hashing network is chosen, $|\Theta|$ is $O(D)$, where $D$ is the data dimension. Traditional algorithms usually use multiple parameters to model one dimension of data\cite{aouad2009grid}\cite{2020Local}, but with convolutional-pooling layers, the number of parameters in a network can be much smaller than the data dimension. On the other hand, $num_m$ can be controlled by setting the length of hash codes $L$ because its upper bound is $2^L$. By selecting proper $\lambda$ in the loss function (\ref{eq3}), we can encourage the hash function to map data to more kinds of hash codes to make the best use of the $L$ bits. Thus $L$ can be set with a small value without unacceptable performance loss, and this part of cost can be limited. Therefore, an upper bound of the total cost is:
 \begin{equation}\label{eq10}
    \hat{C} = 32(2N+1)M|\Theta| + \sum_{m=1}^M (32 + L)2^L (bit).
\end{equation}
Except for some fixed constants, the transmission cost is $O(D)$.

\section{EXPERIMENTS}
This section presents the experiments to evaluate the proposed HDBC algorithm.
\subsection{EVALUATION}
Two measures are chosen to test the quality of clustering results: purity\cite{purity} and normalized mutual information (NMI)\cite{nmi}. The purity counts the number of correct samples divided by the number of total number. For example, given a dataset with size $S$, assume that $\{\mathcal{C}_1,\mathcal{C}_2, ..., \mathcal{C}_I\}$ and $\{\mathcal{T}_1, \mathcal{T}_2, ..., \mathcal{T}_J\}$ are set of clusters and goundtruth respectively. The purity can be calculated by:
\begin{equation}
    purity = \frac{1}{S}\sum_{i=1}^{I} \max_{1\leq j \leq J}{|\mathcal{C}_i\cap\mathcal{T}_j|}.
\end{equation}
Since there is a drawback of purity when the number of clusters is small, the purity cannot evaluate the quality of clustering appropriately. An extreme example is that if all samples are grouped into one cluster, the purity will reach $1$. Thus, the NMI is also included to rule out situations like this and make performance evaluation more comprehensive. The NMI is a variant of a basic concept in information theory called mutual information. Given a set of size $S$, similarly assume that there are $I$ clusters and $J$ true clusters. Use $S_i$, $S^j$ and $S_i^j$ to represent the number of samples in cluster $i$, cluster $j$ and both of them. The NMI can be calculated by:
\begin{equation}
    NMI = \frac{\sum_{i=1}^I\sum_{j=1}^J\frac{S_i^j}{S}\log\frac{SS_i^j}{S_iS^j}}{\sum_{i=1}^I\frac{S_i}{S}\log\frac{S_i}{S}\sum_{j=1}^J\frac{S_j}{S}\log\frac{S_j}{S}}
\end{equation}

In addition, to evaluate the economy of methods, the data volume needed to be transmitted between sub-sites and global site is also considered as cost in the experiments.

\subsection{Datasets}
All algorithms are tested on both synthetic datasets and real-world datasets. Following subsections demonstrate their details.

\subsubsection{Synthetic Datasets}
Since it is convenient to generate a series of datasets with arbitrary numbers of samples, clusters and features, the synthetic datasets are used to test the algorithms when processing HD datasets with big volumes and over hundreds of clusters. Motivated by HD clusters are usually derived from embedding sub-spaces with relatively low dimensions, we follow the method in \cite{2020Local} to generate Gaussian clusters. The explicit procedures to generate an $M$-dimensional Gaussian cluster from a $D$-dimensional embedding sub-space are present as follows:

\begin{itemize}
    \item[1)] Generate a Gaussian matrix $T\in\mathbbm{R}^{M \times D}$. The elements of the matrix are sampled from the normal Gaussian distribution randomly. And then the matrix $T$ is normalized by $\frac{T}{\Vert T \Vert_F}$.
    \item[2)] Generate a shift vector $\textbf{u} \in \mathbbm{R}^{M}$. The elements are sampled from a uniform distribution over $\left[-\frac{20}{log M}, \frac{20}{log M} \right]$. 
    \item[3)] A sample is generated by:
    \begin{equation}
        \textbf{x} = T\textbf{z} + \textbf{u} + \textbf{e}
    \end{equation}
    where $\textbf{z} \sim G(\textbf{0}, \textbf{I}_D)$ and $\textbf{e} \sim G(\textbf{0}, \frac{1}{10M}\textbf{I}_D)$.
    \item[4)] Repeat 3) until generating enough samples to constitute a cluster.
\end{itemize}
We increase the size, dimension, and cluster number of datasets gradually to test how the performances of methods change when datasets become more complex. 
\begin{table}[ht]
    \label{t2}
    \centering
    \caption{Synthetic Datasets, where $s$, $c$, $d$ donate the number of samples, clusters and dimensions, respectively}
    \begin{tabular}{ccccc}
    \toprule
       Dataset  &  $s$ & $c$ & $d(log_2)$ & Embedded Dimensions$(log_2)$ \\
    \midrule
         S1 & $300,000$ & $120$ & $6$ & $1\quad to\quad6$\\
         S2 & $350,000$ & $140$ & $7$ & $1\quad to\quad7$\\
         S3 & $400,000$ & $160$ & $8$ & $1\quad to\quad8$\\
         S4 & $450,000$ & $180$ & $9$ & $1\quad to\quad9$\\
         S5 & $500,000$ & $200$ & $10$ & $1\quad to\quad10$\\
    \bottomrule
    \end{tabular}
    \begin{tablenotes}
        \item The embedded dimensions donate the parameter $D$ in the foregoing data-generate procedures. For each embedded dimension, $20$ different clusters are generated.
    \end{tablenotes}
\end{table}

\subsubsection{Real-world Datasets}
Eight real-world datasets are selected for the experiments, including three hyperspectral datasets Salinas, Pavia Centre, Pavia University\cite{hyper}; four datasets from UCI\cite{Dua:2019}; and handwritten digits MNIST\cite{MNIST}. The summary of eight datasets is presented in Table 2.
\begin{table}[ht]
    \label{t3}
    \centering
    \caption{Eight Real-World Datasets,\\ where $s$, $c$, $d$ are defined in the same way as TABLE 1}
    \begin{tabular}{cccc}
    \toprule
       Dataset  &  $s$ & $c$ & $d$ \\
    \midrule
         Salinas & $111,104$ & $16$ & $204$ \\
         Pavia Centre & $783,640$ & $9$ & $102$ \\
         Pavia University & $206,400$ & $9$ & $103$ \\
         WBCD & $569$ & $2$ & $30$ \\
         Waveform-noise & $5,000$ & $3$ & $40$ \\
         LSD & $6,435$ & $6$ & $36$ \\
         WFRND & $5,456$ & $3$ & $24$ \\
         MINST & $60,000$ & $10$ & $784$ \\
    \bottomrule
    \end{tabular}
\end{table}

Some statistics and descriptions of the datasets are presented as follows.
\begin{itemize}
    \item Salinas was collected by AVIRIS sensor over Salinas Valley. The scene contains $512 \times 217$ pixels, and each pixel is featured by 224 bands. But 20 water absorption bands are discarded, so the dimension is actually 204. Salinas contains 16 classes. 
    \item Pavia Centre and University scenes are acquired by ROSIS sensor over Pavia, Italy. After discarding some pixels with no information, the images for analysis are $1096 \times 715$ for Pavia Centre and $610 \times 340$ for Pavia University. Pavia Centre contains $102$ bands and Pavia University contains $103$ bands. Both of them include $9$ classes. 
    \item Wisconsin breast cancer dataset (WBCD) is from UCI. The features are computed from a digitized image of a fine needle aspirate (FNA) of a breast mass. There are two classes representing malignant and benign, respectively.
    \item Waveform-noise from UCI is constructed by three classes of waves and each of them is generated from a combination of three base waves. All forty features include noise, and the latter nineteen are totally noise with mean $0$ and variance $1$.
    \item Landsat satellite dataset (LSD) from UCI is generated from data purchased from NASA. The dataset consists of the multi-spectral values of pixels in $3 \times 3$ neighborhoods in a satellite image and the classification associated with the central pixel in each neighborhood.
    \item Wall-following robot navigation dataset (WFRND) from UCi is collected by ultrasound sensors as the SCITOS G5 robot navigates through the room, following the wall in a clockwise direction for 4 rounds. The dataset contains three classes representing the movements of the robot. 
    \item MNIST consists of $60,000$ samples of handwriting digits. All samples are 784-dimensional vectors ($28\times 28$ images) valued by grayscales. It contains $10$ classes from 0 to 9. 
\end{itemize}


\subsection{Comparison with Centralized Clustering}

\begin{figure*}[b]
    \centering
    \includegraphics{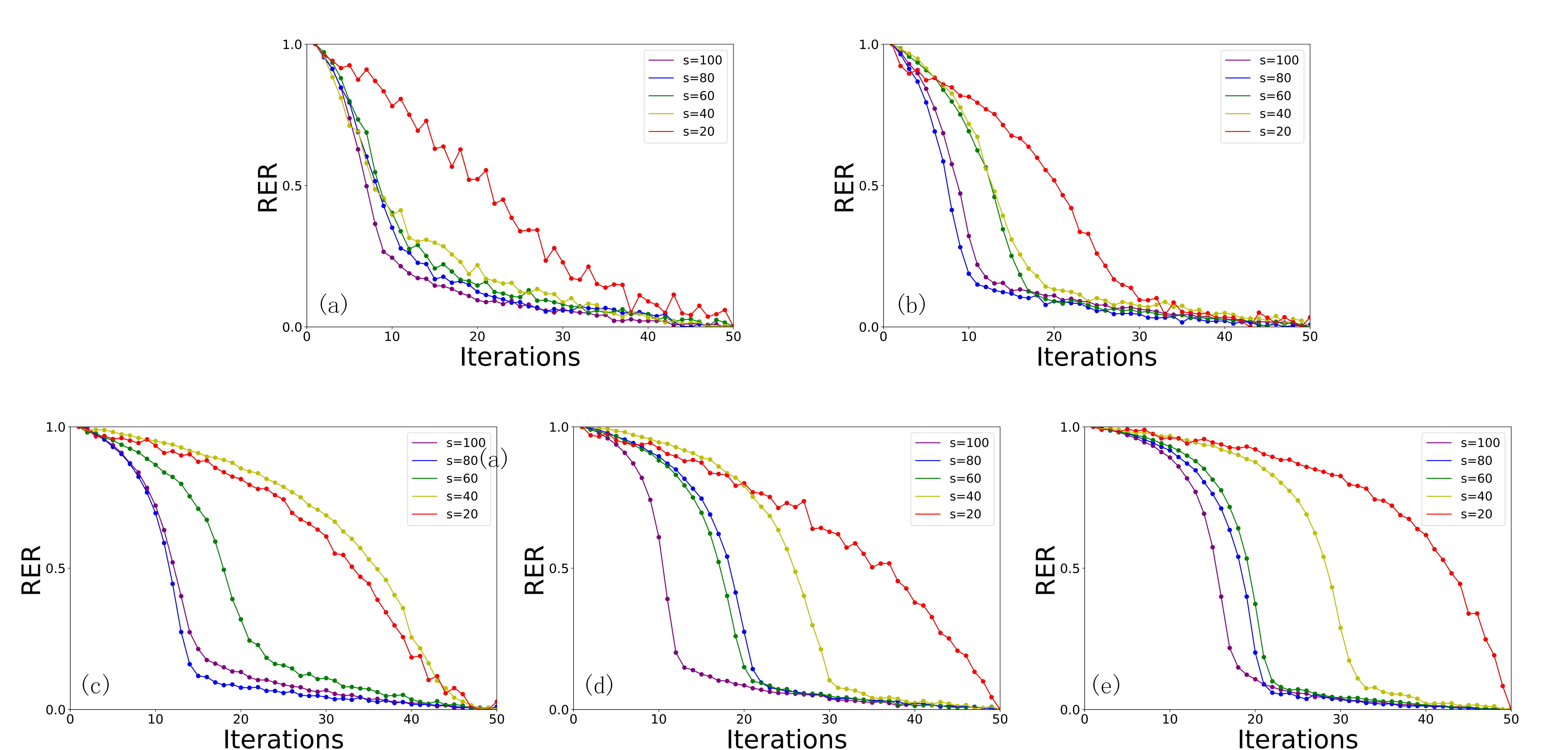}
    \caption{Convergence analysis of the HBDC with different numbers of sub-sites, donated by $s$, on five synthetic datasets: (a) S1, (b) S2, (c) S3, (d) S4, (e) S5.}
    \label{fig:5}
\end{figure*}

To test the representative capability of hash codes, we compare the HBDC in the distributed situation with centralized clustering algorithms to test how representative the hash codes can be. In this subsection, we conduct experiments on four datasets: WBCD, Waveform-noise, LSD and WFRND, because executing centralized clustering on big HD datasets in one machine could be very time-consuming and these four datasets are relatively small. Since spectral clustering is the downstream clustering method for HBDC, it is chosen as the centralized clustering method for comparison intuitively. In addition, we also use k-means for supplementary when spectral clustering cannot perform well.

\begin{table}[t]
\label{T3}
\centering
\caption{Performance of Centralized Clustering and distributed clustering HBDC in Terms of Purity(\%) and NMI(\%).} 
\resizebox{9cm}{!}{
 \begin{tabular}{ccccccc}
 \toprule
  & \multicolumn{2}{c}{Spectral} & \multicolumn{2}{c}{K-means} & \multicolumn{2}{c}{HBDC}\\
  \cmidrule(r){2-3}  \cmidrule(r){4-5} \cmidrule(r){6-7}
 Datasets & Purity & NMI & Purity & NMI & Purity & NMI\\ 
   \midrule
 WBCD & $86.29$ & $49.87$  & $92.79$& $62.30$& $86.29$& $40.60$\\ 
  Waveform-noise & $51.42$ & $36.80$  & $52.04$& $36.54$& $53.48$& $36.44$\\  
  LSD & $63.41$ & $50.28$  & $74.17$& $61.23$& $69.12$& $53.56$\\ 
WFRND & $43.67$ & $11.36$  & $49.63$& $11.08$& $62.93$& $12.07$\\ 
   \bottomrule
 \end{tabular}}
 \begin{tablenotes}
        \item The cluster numbers in all three methods are set as real class numbers. The parameter $\gamma$ in spectral clustering is $1$. In the HBDC, the length of hash codes is 4bit for WBDC, 4bit for waveform-noise, 12bit for LSD and 8bit for WFRND.
    \end{tablenotes}
\end{table}

In order to simulate the distributed situation, each dataset is distributed  into ten sub-sites randomly, and the size of data at each site is larger than fifty to make sure all sites have enough data for local training. It is not difficult to extend to a more general case in reality. Even though under extreme circumstances the data in some sub-sites are very small, we can reduce their batch sizes and make the equation (\ref{eq4}) weighted average. Besides, the structure of hashing network in this part is set as three fully connected layers. The numbers of units in the first two layers are equal to the dimension of input data and use $Relu$ as active functions. The output layer is of the length of hash codesthe  with $tanh$ as an active function. This $tanh$ is replaced by $sign$ after the training part to generate binary codes. 

Table 3 shows the clustering performance in terms of purity and NMI. On the later three datasets, the HDBC can outperform centralized spectral clustering. This is because the original spectral clustering is unsuitable for processing HD data. As the dimension of space grows, the distribution of samples becomes sparse and their connection becomes weak, so edge cutting may not lead to promising results. By contrast, graphs generated by shorter hash codes have edges whose weights vary with a greater scale, so the input adjacent matrix is denser. On the other hand, the weak performance on the WBCD may be attributed to the small size of the dataset. Whereas it still achieves eighty percent of the centralized case. On the WBCD and LSD, the results of k-means are better than spectral clustering. The reason is the differences between the numbers of samples in different classes are relatively large, but spectral clustering based on (\ref{eq6}) tends to group samples evenly.

\subsection{Convergence}
The convergence of distributed training is a substantial factor in the HBDC. We report it on five synthetic Gaussian datasets to see the convergence rate and its sensitivity to the number of sub-sites. In this subsection, the convergence is represented by the relative error ratio (RER), which is defined as:
\begin{equation}
    RER = \frac{\mathcal{L}-\mathcal{L}_{min}}{\mathcal{L}_{max}-\mathcal{L}_{min}}
\end{equation}
where $\mathcal{L}$, $\mathcal{L}_{min}$ and $\mathcal{L}_{max}$ donate the value, minimum and maximum of the loss function (\ref{eq2}), respectively.

Fig \ref{fig:5} shows how the RER changes as the number of iterations grows. It is clear that the number of sub-sites influences the rate of convergence significantly. In general, more sub-sites result in faster convergence. The reason is that with more sub-sites, more samples participate in training within certain iterations, so it influences the convergence in the same way as increasing batch size. But there are still differences on different datasets. For example, S5 is the most complex dataset, so subfigure (e) shows the slowest descent of the curves. Even with $100$ sub-sites, the model still takes nearly $20$ iterations for convergence, It is similar on S4 that curves representing more sub-sites fall faster. The model with $100$ sub-sites almost converges within $10$ iterations, which is faster than that on S5. Whereas, when datasets become more simple, in addition to the faster overall convergence rate, the acceleration brought by more sub-sites is not absolute. The differences between the curves become smaller gradually from (e) to (a). On the relatively simple datasets S2 and S3, the blue curves (representing the model with $80$ sub-sites) descend faster than the purple curves (representing the model with $100$ sub-sites). When it comes to S1, the model with $40$ sub-sites can converge at a high rate, and keeping increasing the number $s$ cannot bring more conspicuous benefit.

\subsection{Horizontal Comparison}
In this subsection, five state-of-the-art distributed clustering algorithms are selected deliberately, including a partition-based algorithm k-means$||$\cite{bahmani2012scalable}, two density-based algorithms DBDC\cite{2004DBDC} and LSHDDP\cite{zhang2016efficient}, and two model-based algorithms LDSDC\cite{2020Local} and REMOLD\cite{liang2017remold}. We report the methods on all synthetic datasets and four HD real-world datasets: Salinas, Pavia Centre, Pavia University and MNIST.
\subsubsection{Settings of Methods}
For the best performance, the parameters of each method are searched by multiple times of experiments. The settings of the baseline algorithms and their description are presented as follows:
\begin{itemize}
    \item HBDC's hashing networks are designed according to the dimensions of the datasets. The convolutional-pooling layer is used to process HD data like synthetic Gaussian datasets and MNIST because the pooling layer can exponentially decrease the dimensions of hidden features and the number of parameters in the hashing network. We also use hyperspectral datasets to test the performance of fully-connected hashing networks. Besides, $\lambda$, $ t $ are selected in the set $\{ 0.1, 0.2, 0.5, 1, 2, 5\}$. The number of iterations for training $n$ is in $\{ 5, 10, 15, 20 \}$. 
    \item k-means$||$'s only parameter $k$ represents the number of clusters. We set $k$ equal to the number of real clusters in each dataset. The initial seeds are randomly chosen by k-means++.
    \item DBDC uses $\epsilon$ cut-off density, where $\epsilon$ is estimated as the second percentile of the ascending ordered distances of all sample pairs according to \cite{zhang2016efficient}. Another parameter $\lambda$ is in $\{0.01, 0.05, 0.1, 0.5, 1\}$.
    \item LSHDDP also uses $\epsilon$ cut-off density. Following  \cite{zhang2016efficient}, we set $M=10$ and $\pi=3$. The number of clusters $k$ is set as the real number. The samples with top $k$ gamma values are the center of $k$ clusters. 
    \item LDSDC follows the setting in \cite{2020Local}. The parameter $D$ is set to the smallest integer such that the retained eigenvalue ratio of sample covariance matrix is greater than $70\%$. Set $K=\sqrt{N/W}$, where $N$ and $W$ represent the numbers of samples and sub-sites, respectively.
    \item REMOLD's parameter $K$ is set in the same way as LDSDC. The candidate set $\Delta$ for parameter $\delta$ is obtained by an auxiliary algorithm according to \cite{liang2017remold}.
\end{itemize}

\subsubsection{Experimental Results}
\begin{figure}[ht]
    \centering
    \includegraphics{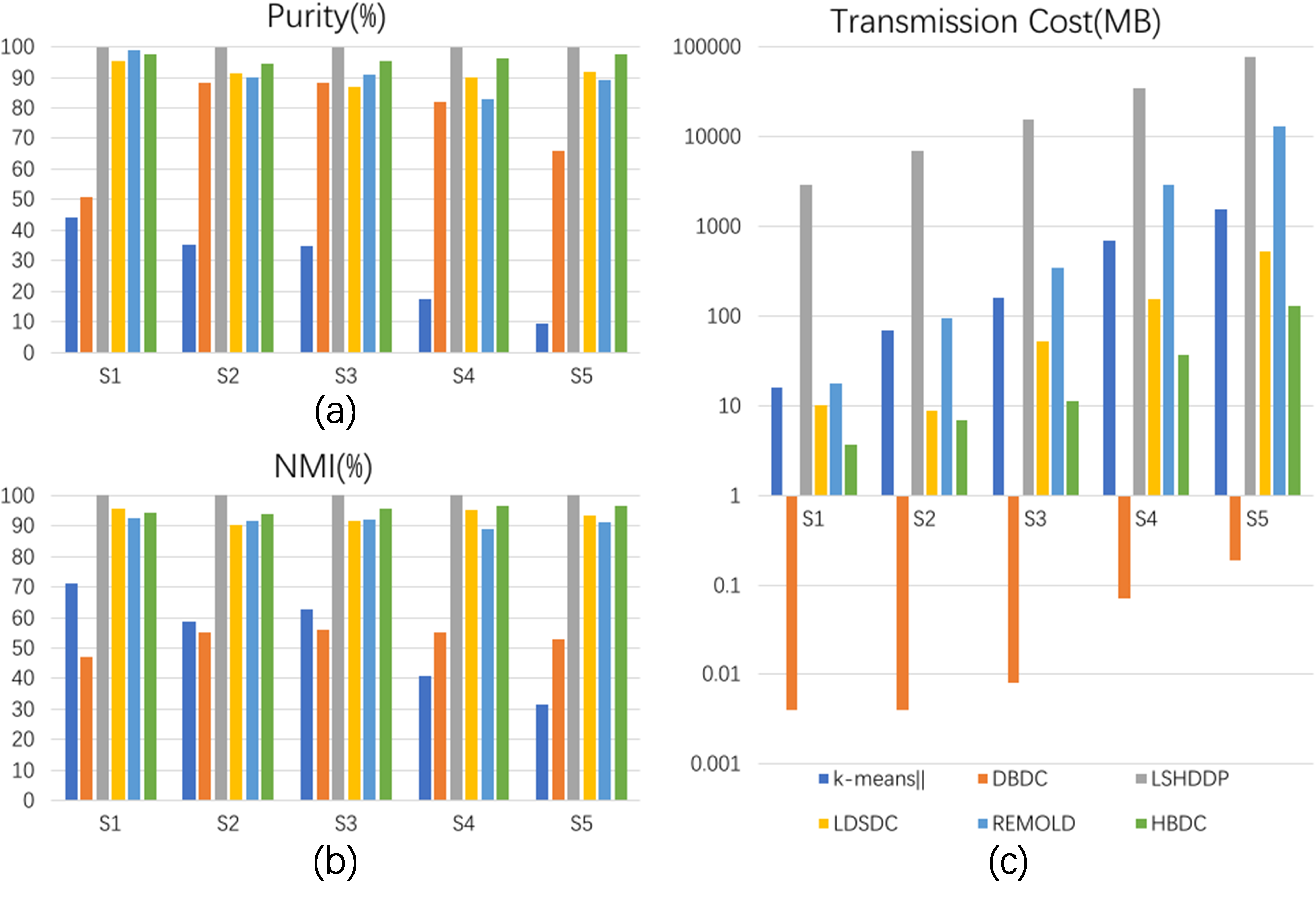}
    \caption{Experimental results on five synthetic datasets. (a) purity, (b) NMI, (c) transmission cost, respectively. The distributed situation is simulated by $1$ global site and $20$ sub-sites. The structure of the hashing network in HBDC consists of two convolutional layers followed by a fully connected layer that maps data into $64$-bit hash codes.}
    \label{fig:6}
\end{figure}

The performance of clustering results on the synthetic datasets S1 to S5 is shown in Fig \ref{fig:6}. From (a) and (b), it can be seen that LSHDDP (represented by gray bar) achieves perfect performance in terms of accuracy. HBDC (represented by green bar) also performs well and achieves more than $95\%$ purity and NMI scores except for S2, and this has not been influenced by the increase of the dimension and volume of datasets. Its purity and NMI are stable on all five datasets. With the usage of hashing networks, HBDC is more capable of processing HD data and shows better robustness. LDSDC (represented by yellow bar) and REMOLD (represented by light blue bar) show comparable and stable performance, which is almost more than $90\%$ purity and NMI measures. By contrast, the performance of k-means$||$ (represented by dark blue bar) gradually degraded as the index of the dataset increases. This can be attributed by that its local k-means clustering cannot model local data in sub-sites accurately. When dataset becomes more complex, the loss of information also rises so that it is not sufficient enough to yield a reliable result. Likewise, DBDC (represented by orange bar) shows low-quality performance, although its purity scores on S2, S3 and S4 are relatively high because it divides most samples into a few groups. The reason is similar to that of k-means$||$, which is that a simplified model results in too much information loss when processing complex data.

The transmission cost is shown in (c) in terms of megabytes by semi-log scale.  Generally, the cost of all algorithms rises gradually as the index of the dataset because of the linear increase of data dimension. LSHDDP shows the highest cost because it needs to reshuffle dataset multiple times. K-means$||$ also costs high for data transmission because its cost is proportional to both the data dimension and class number. In the two model-based methods, as a modified version, LDSDC costs much lower than REMOLD, but it still needs $152.30$MB and $520.82$MB for HD datasets S4 and S5, while HBDC only transmits $36.71$MB and $129.58$MB, including uplink and downlink. The cost reduction is more obvious on HD datasets than on LD datasets. This benefits from the efficiency of binary hash code so that we do not need to transmit HD vectors. The cost for DBDC is the lowest, because it does not need to transmit more data even though the size of dataset is large.

\begin{figure}[ht]
    \centering
    \includegraphics{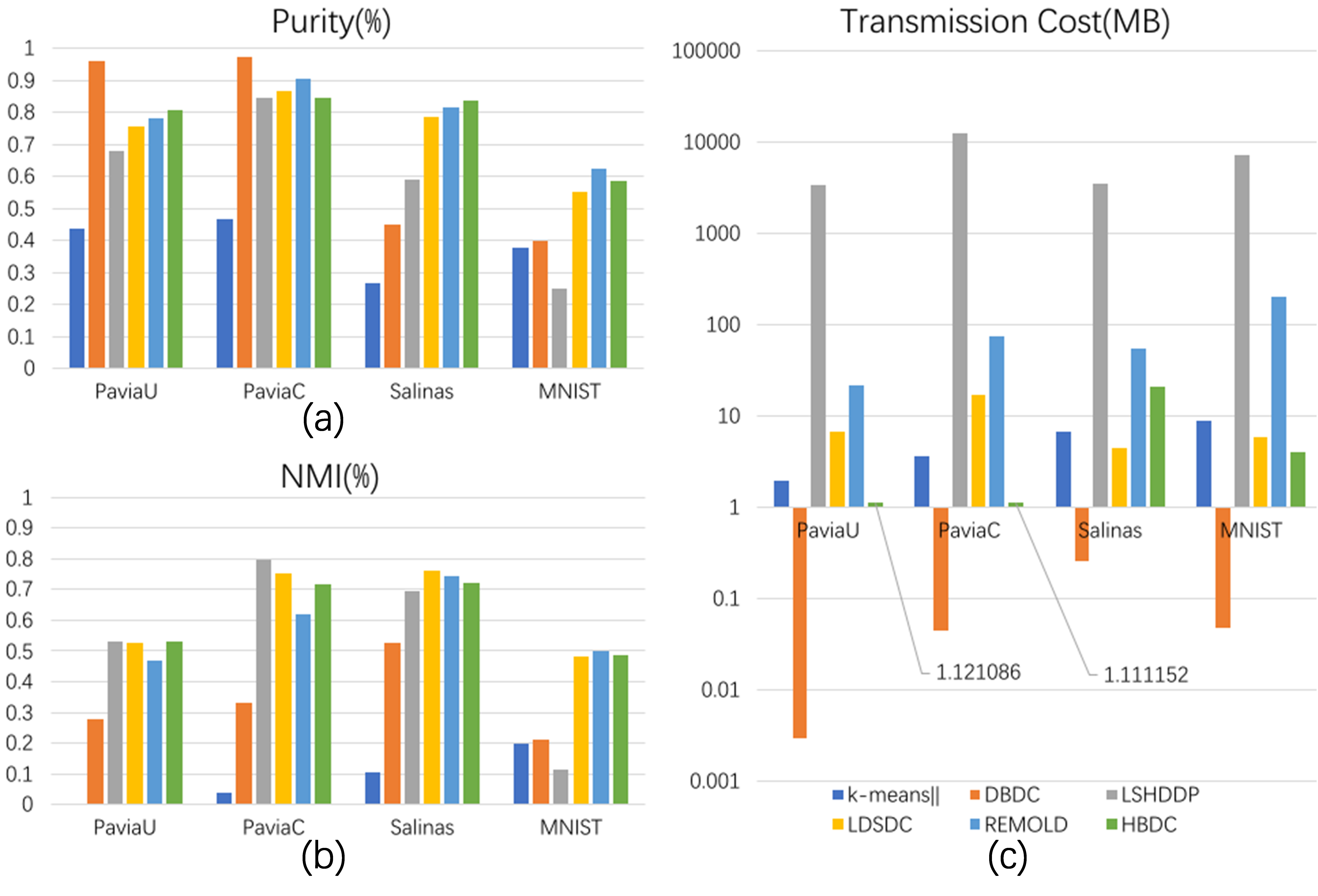}
    \caption{Experimental results on five real-world datasets. (a) purity, (b) NMI, (c) transmission cost, respectively. The distributed situation is simulated by $1$ global site and $10$ sub-sites. HBDC uses 3-layer fully connected networks for hyperspectral datasets and a 2-layer convolutional network followed by a linear output layer for MNIST. The hash codes are 8-bit for Pavia Centre and University, 24-bit for Salinas and 16-bit for MNIST.}
    \label{fig:7}
\end{figure}

Fig \ref{fig:7} shows the clustering performance on four real-world datasets. In (a), DBDC has the highest purity score on PaviaU and PaviaC, but that sharply drops on Salinas and MNIST. LDSDC, REMOLD and HBDC show similar performance. In specific, REMOLD achieves the highest purity score on PaviaC and MNIST, while HBDC outperforms on PaviaU and Salinas. Comparing (a) and (b), the large difference between purity and NMI usually means unbalanced clustering results, such as k-means$||$ and DBDC on PaviaC and PaviaU. LSHDDP performs well on hyperspectral datasets but shows the lowest purity and NMI on MNIST. Considering the purity and NMI simultaneously, HBDC is either the best or the second which is close to the first. 

In (c), DBDC and LSHDDP still show the lowest and the highest cost. It can be observed that among the three most robust algorithms, HBDC has the lowest transmission cost except for Salinas. The exception on Salinas is caused by the heavy fully-connected structure of the hashing network and its high dimension, which indicates that convolutional networks are more suitable for HD data. This can also be demonstrated by the performance on MNIST: the cost is $4.00$MB for HBDC, $5.79$MB for LDSDC and $202.82$MB for REMOLD although its dimension is much higher than Salinas. Overall, the results on five synthetic datasets and four real-world datasets illustrate the superiority of the proposed HBDC by comparing clustering accuracy, robustness and cost with other state-of-the-art algorithms.

\section{Conclusion}
Motivated by the outstanding performance of the learning-to-hash technique in the nearest neighbor searching, we propose a hashing-based distributed clustering (HDBC) algorithm. With the usage of hashing networks, The proposed algorithm is capable of clustering massive HD data at low cost and preserving privacy. HBDC exploits a learnable hashing network as its hash function, and the training of hashing network consists of global model merging, broadcasting and local training, which is driven in a self-supervised manner and accelerated by a sample selection mechanism. Original data are mapped into hash codes by hash function and form a graph for final graph-cutting clustering. Some experiments on both synthetic and real-world datasets demonstrate that HBDC has better comprehensive performance than the other four benchmark algorithms by weighting accuracy, robustness and cost. In addition, all the experiments are conducted with randomly initialized networks, which causes difficulty for the training, while much research on hashing methods uses pre-trained models. Since pre-trained models are gaining more popularity, HBDC has more possibilities and higher limits. This can be a feasible direction for future research.

Finally, HBDC can be easily extended to an online-clustering version. When the local data in sub-sites change gradually or sub-sites generate new data, the hash network can be adjusted in real-time by distributed training to adapt to new data distribution. This can be applied to many real cases like communication and industrial manufacturing.


%





\ifCLASSOPTIONcaptionsoff
  \newpage
\fi



%

\bibliographystyle{IEEEtran}
\bibliography{reference.bib}




%

\begin{IEEEbiography}
[{\includegraphics[width=1in,height=1.25in,clip,keepaspectratio]{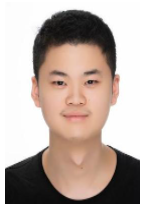}}]{Yifeng Xiao}
received the B.S. degree in applied
mathematics from Xi’an Jiaotong University, Xi’an,
China, in 2021, where he is currently pursuing the
master's degree in applied mathematics. His current
research interests include the machine learning, distributed algorithm and learning to optimize.
\end{IEEEbiography}

\begin{IEEEbiography}
[{\includegraphics[width=1in,height=1.25in,clip,keepaspectratio]{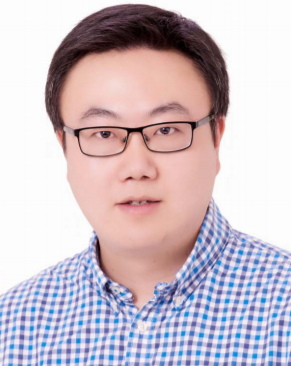}}]{Jiang Xue}
(Senior Member, IEEE) received the
B.S. degree in information and computing science
from the Xi’an Jiaotong University, Xi’an, China,
in 2005, the M.S. degree in applied mathematics
from Lanzhou University, China, and Uppsala University, Sweden, in 2008 and 2009, respectively, and
the Ph.D. degree in electrical and electronic engineering from ECIT, Queen’s University of Belfast,
U.K., in 2012. From 2013 to 2017, he was a
Research Fellow with the University of Edinburgh,
U.K. Since 2017, he has been with the National
Engineering Laboratory for Big Data Analytics, Xi’an International Academy
for Mathematics and Mathematical Technology, School of Mathematics and
Statistics, Xi’an Jiaotong University, Pengcheng Laboratory, China. He is supported by the Zhongying Young Scholars Project. His main interests include
the machine learning and wireless communication, performance analysis of
general multiple antenna systems, CSI estimation and prediction, stochastic
geometry, cooperative communications, and cognitive radio.
\end{IEEEbiography}


\begin{IEEEbiography}
[{\includegraphics[width=1in,height=1.25in,clip,keepaspectratio]{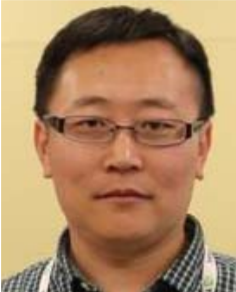}}]{Deyu Meng}
received the B.Sc., M.Sc., and Ph.D. degrees from Xi’an Jiaotong University, Xi’an, China, in 2001, 2004, and 2008, respectively. He was a Visiting Scholar at Carnegie Mellon University, Pittsburgh, PA, USA, from 2012 to 2014. He is currently a Professor with the School of Mathematics and Statistics, Xi’an Jiaotong University, and an Adjunct Professor with the Faculty of Information Technology, Macau University of Science and Technology, Taipa, Macau, China. His research interests include model-based deep learning, variational networks, and meta learning.
\end{IEEEbiography}




\end{document}